\documentclass[24pt]{article}
\usepackage{amssymb}
\usepackage{amsfonts}
\usepackage{amssymb,amsmath}
\usepackage{mathrsfs}
\usepackage{latexsym}
\usepackage{amsmath}
\usepackage{latexsym}
\usepackage[cp1251]{inputenc}
\usepackage{graphicx}
\usepackage{color}

\title{Application of a quantum wave impedance approach for a quantum capacitance calculation}
\author{O. I. Hryhorchak\\
{\small Department for Theoretical Physics, Ivan Franko National
University of Lviv,}\\
{\small 12, Drahomanov Str., Lviv, UA--79005,
Ukraine}\\
\small{\it{Orest.Hryhorchak@lnu.edu.ua}}}

\def\ch{\mathop{\rm ch}\nolimits}
\def\sh{\mathop{\rm sh}\nolimits}
\def\th{\mathop{\rm th}\nolimits}

\def\tg{\mathop{\rm tg}\nolimits}

\begin{document}
\renewcommand{\abstractname}{Abstract}
\maketitle

\begin{abstract}
A method for a calculation of quantum capacitance for a two-dimesional electron gas (2DEG) in potential wells of complicated geometry on the base of a quantum wave impedance technique was proposed. The application of this method was illustated on four different forms of potential wells: infinite and finit rectangular well, a finit rectangular double well and a parabolic double well.
It was shown how a quantum capacitance of these wells depends on its parametres. Numerical calculations of a quantum capacitance for mentioned potential wells were made at temperature $T=0$ and the results were presented graphically.
\end{abstract}

\section{Introduction}
We often talk about a nanoscopic scale as a point which signals that the properties of materials change. So from the one side of this point we have a subatomic level for describing of which we directly use the Shr\"{o}dinger equation and from the other side we have the ``bulk'' level where the ``volume'' effects significantly dominate on the individual properties of each small component. To describe this level we use methods of statistical physics where the thermodynamic limit is applicable. On the nanoscale level both these worlds meet each other and therefore completely unique phenomena arise, where both the fluctuations in the averaged properties and geometry
of the material begin to play a significant role in the behavior and properties of a system, and we should take it into account. This uniqueness of a nanoworld allows one to create miniature prototypes of macroscopic objects, which one can call nanorobots. In particular in the paper \cite{MacKinnon:2005} two simple nanomodels, namely quantum gears and a quantum shuttle, are considered and they (according to the author's statement) serve as model systems for a wide range of other nanodevices. Another promising area of a nanotechnology is the creation of quantum logic elements \cite{Bertoni_etall:2000}.

One of the phenomena which appeare at the interface of these two worlds is a quantum capacitance. For the first time this  was introduced by Serge Luryi \cite{Luryi:1988} in 1988 to characterize the changing of a chemical potential in low-dimentional quan\-tum systems. In this article he calculated a quantum capacity for a 2DEG (two-dimentional electron gas). Three years later Luryi applied this concept for studying resonant tunnelling diodes \cite{Luryi:1991}. In \cite{Rahman_etall:2003, Burke:2003} a quantum capacitance term was used in the modeling of carbon nanotube devices. Article \cite{Johna_Castro_Pulfreyb:2004}  is dedicated to a
quantum capacitance in nanoscale device modeling
The authors concluded that  for fully turned-on devices the models must include quantum capacitance in order to
properly capture the device behavior. The also revealed  the relationship between the transconductance of a carbon nanotube field-effect transistor (CNFET) and a quntum capacitance .

In this paper we invetigate only systems in which the discretization of energy levels is caused by limitation of potential energy in one dimension, so in fact we deal with a 2DEG. We consider four different forms of a potential energy which appear as a result of limitation in one dimension. For energy levels calculation we use a quantum impedance method which demonstarated its efficasy for investigation of quantum mechanical systems with complicated geometry of a potential \cite{Kabir_Khan_Alam:1991, Nelin_Imamov:2010, Babushkin_Nelin:2011_1, Khatyan_Gindikina_Nelin:2015, Ashby:2016, Hague_Khondker:1998, Nelin:2008, Nelin:2009, Nazarko_etall:2010, Nelin:2011, Nelin:2009_1, Nelin_Sergiyenko:2008, Gindikina_etall:2015, Vodolazka_Mikolaychik_Nelin:2017} 
 
\section{Quantum capacitance for a single well}
In this section we start from the simplest cases, namely the rectangular potential well of width $a$ (in one direction) with finite and infinite walls. In two other perpendicular directions (let these are the directions of the x and y axes) we assume an absence of any edges. First, we consider a rectangular potential well with infinite walls and then with finite ones.

According to the definition we have the following expression for a specific quantum capacity
\begin{eqnarray}
C_q=e^2\frac{\partial n}{\partial \mu},
\end{eqnarray}
where $n$ is a 2D concentration of electrons, $\mu$ is an electrochemical potential and $e$ is a charge of an electron. Then
\begin{eqnarray}
n=\frac{2}{(2\pi)^2}\sum_{j=1}^N 
\int\limits_{-\infty}^{\infty} dk_x \int\limits_{-\infty}^{\infty} dk_y \theta\left(\mu-\frac{\hbar^2(k_x^2+k_y^2)}{2m^*}-E_j\right),
\end{eqnarray}
where $k_x$ and $k_y$ are components of a wave-vector $\vec{k}$ which is in a $x-y$ plane, $m^*$ is an effective mass of an electron, $E_i$ are values of bound states energies of an electron in the considered system. Having made the transition to the polar coordinate system 
\begin{eqnarray} 
n=\frac{1}{\pi}\sum_{j=1}^N \int\limits_0^{\sqrt{2m^*(\mu-E_j)}/\hbar}kdk\theta\left(\mu-\frac{\hbar^2}{2m^*}(k_x^2+k_y^2)-E_i\right)\end{eqnarray}
and after making an integration we get
\begin{eqnarray}\label{n_mu}
n=\sum_{j=1}^N \frac{m^*(\mu-E_j)}{\pi\hbar^2}\theta\left(\mu-E_j\right)
\end{eqnarray} 
and
\begin{eqnarray}\label{Cq_mu}
C_q=e^2\frac{\partial n}{\partial\mu}=\frac{m^*e^2}{\pi\hbar^2}\sum_{j=1}^N \theta\left(\mu-E_j\right).
\end{eqnarray}
For a potential well with infinite walls the expression for $E_j$ is well-known 
\begin{eqnarray}
E(N)=\frac{\hbar^2}{2m}\frac{\pi^2}{a^2}j^2.
\end{eqnarray}
After that using both (\ref{n_mu}) and (\ref{Cq_mu}) we can find the result for $C_q$ numerically. This result is presented on Figure 1.
\begin{figure}[h!]
	\centerline{
		\includegraphics[clip,scale=0.9]{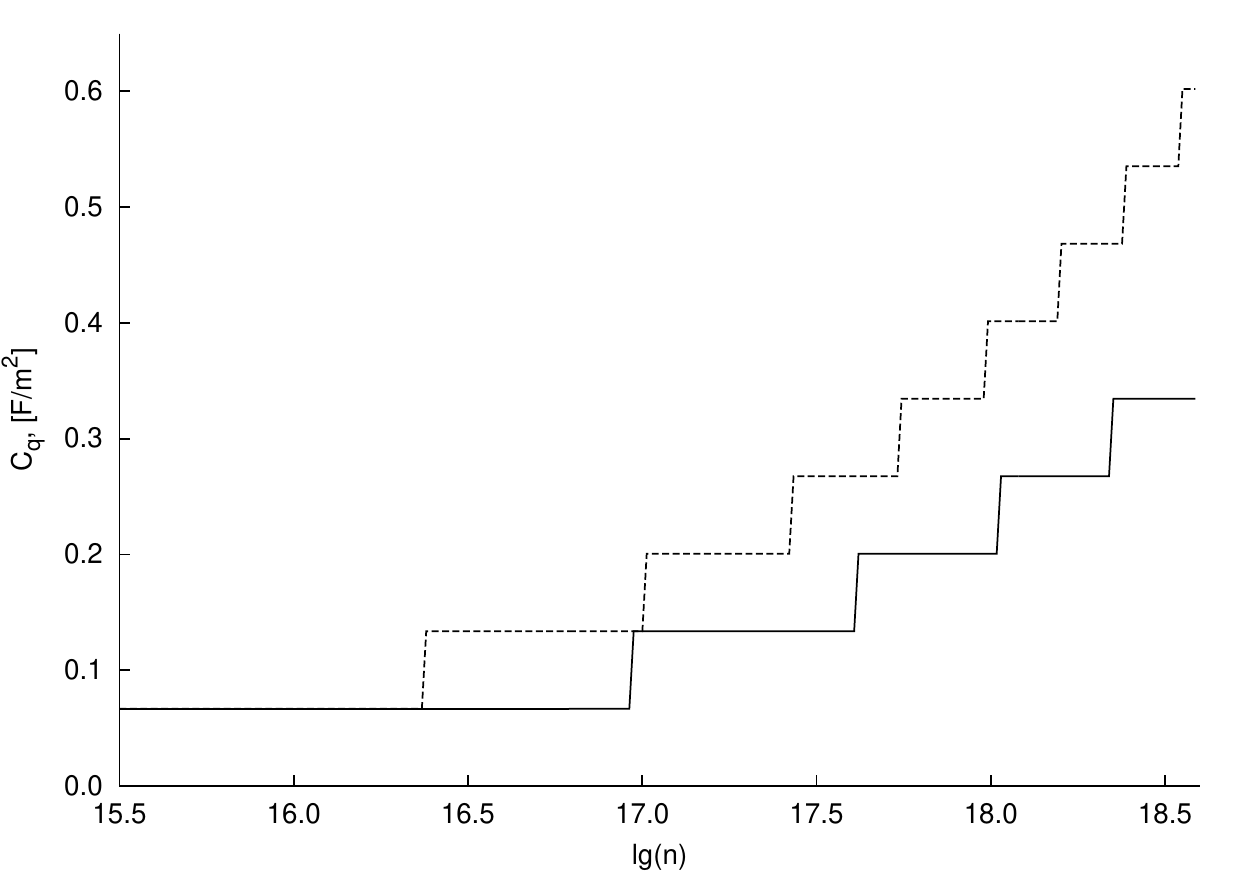}}
	\caption{\small{Dependence of a specific quantum capacity $C_q$ on $lg(n)$ (n is the 2D concentration of electrons) for a 3D system with a potential well (in one direction) of a width $a$ with infinite walls. Solid line is for $a=5$ nm, dashed line is for $a=10$ nm. Effective mass of a particle in both cases is $m^*=0.1m_0$, where $m_0$ is a mass of an electron.}}
	\label{fig:U1U2U3}
\end{figure}

Now we consider a well of a finite depth $U_w>0$ and use an impedance approach to calculate energy levels of this system:
\begin{eqnarray}
-z_0=z_w\frac{z_0\ch[\gamma_w a]-z_w\sh[\gamma_w a]}
{z_w\ch[\gamma_w a]-z_0\sh[\gamma_w a]},
\end{eqnarray}
where
\begin{eqnarray}
z_w=\sqrt{\frac{2(E+U_w)}{m}},\qquad 
\gamma_w=\frac{im}{\hbar}z_w, \qquad z_0=\sqrt{\frac{2E}{m}}.
\end{eqnarray}
It gives
\begin{eqnarray}
\th[\gamma_wa]=\frac{2z_0z_w}{z_w^2+z_0^2}.
\end{eqnarray}

Taking into account expressions for $z_0$, $z_w$ and $\gamma_w$ we get
\begin{eqnarray}
\tg\left(\frac{\sqrt{2m(E+U_w)}a}{\hbar}\right)=
2\frac{\sqrt{-E(U_w+E)}}{U_w},\!\!\
\!\!\! \quad -U_w\leq E \leq 0.
\end{eqnarray}
On the base of this expression and using numerical calculations we find energy levels $E_j$ and then a
quantum capacity $C_q$ on the base of formulas (\ref{n_mu}) and (\ref{Cq_mu}). The results of a numerical calculation are on Figure 2 and Figure 3.

\begin{figure}[h!]
	\vspace{-0.2cm}
	\centerline{
		\includegraphics[clip,scale=0.9]{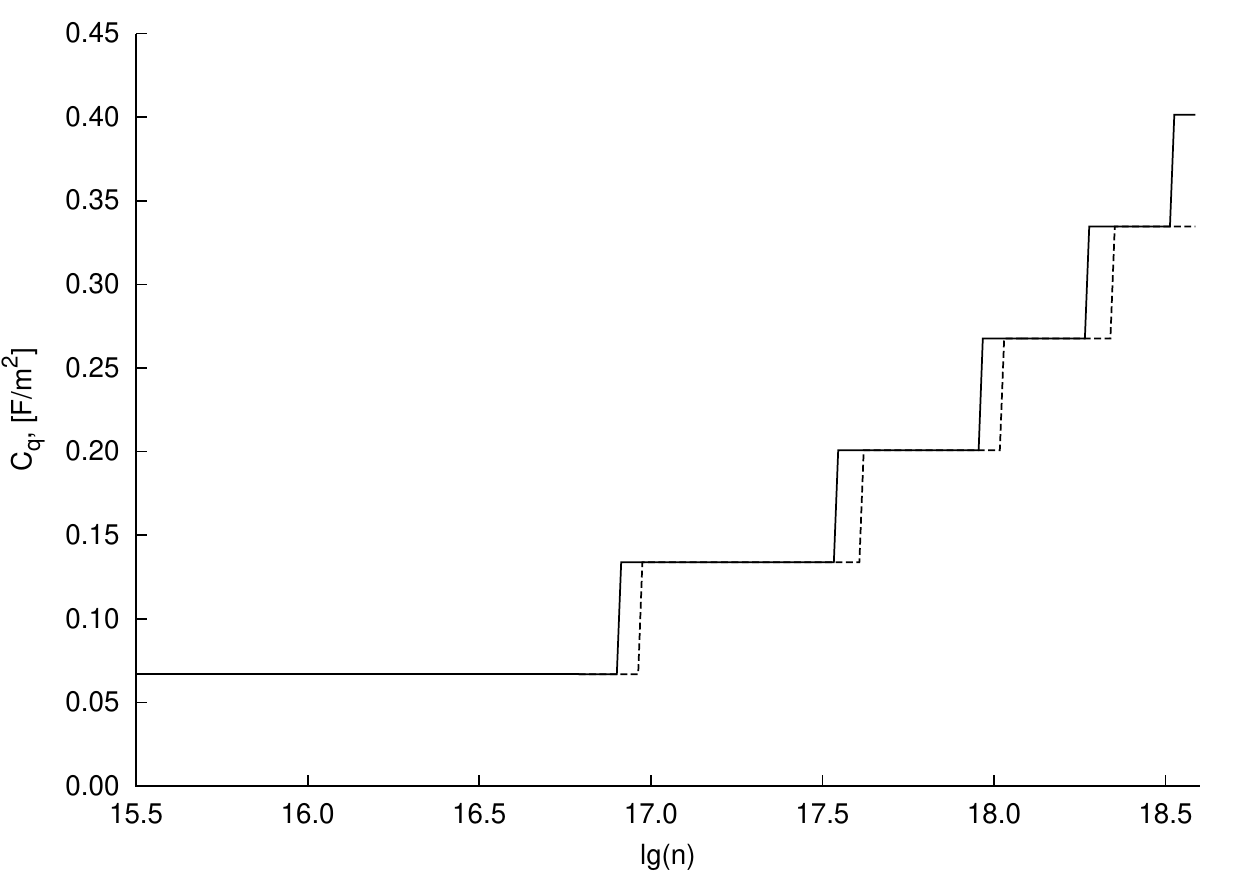}}
	\vspace{-0.3cm}
	\caption{\small{Dependence of a specific quantum capacity $C_q$ on $\lg(n)$ (n is the 2D concentration of electrons) for a 3D system with a potential well (in one direction) of a width $a=5$ nm. Solid line is for well with finit walls, dashed line is for a well with infinite walls. An effective mass of a particle in both cases is $m^*=0.1m_0$, where $m_0$ is a ``bare'' mass of an electron.}}
\end{figure}
\begin{figure}[h!]
	\centerline{
		\includegraphics[clip,scale=0.9]{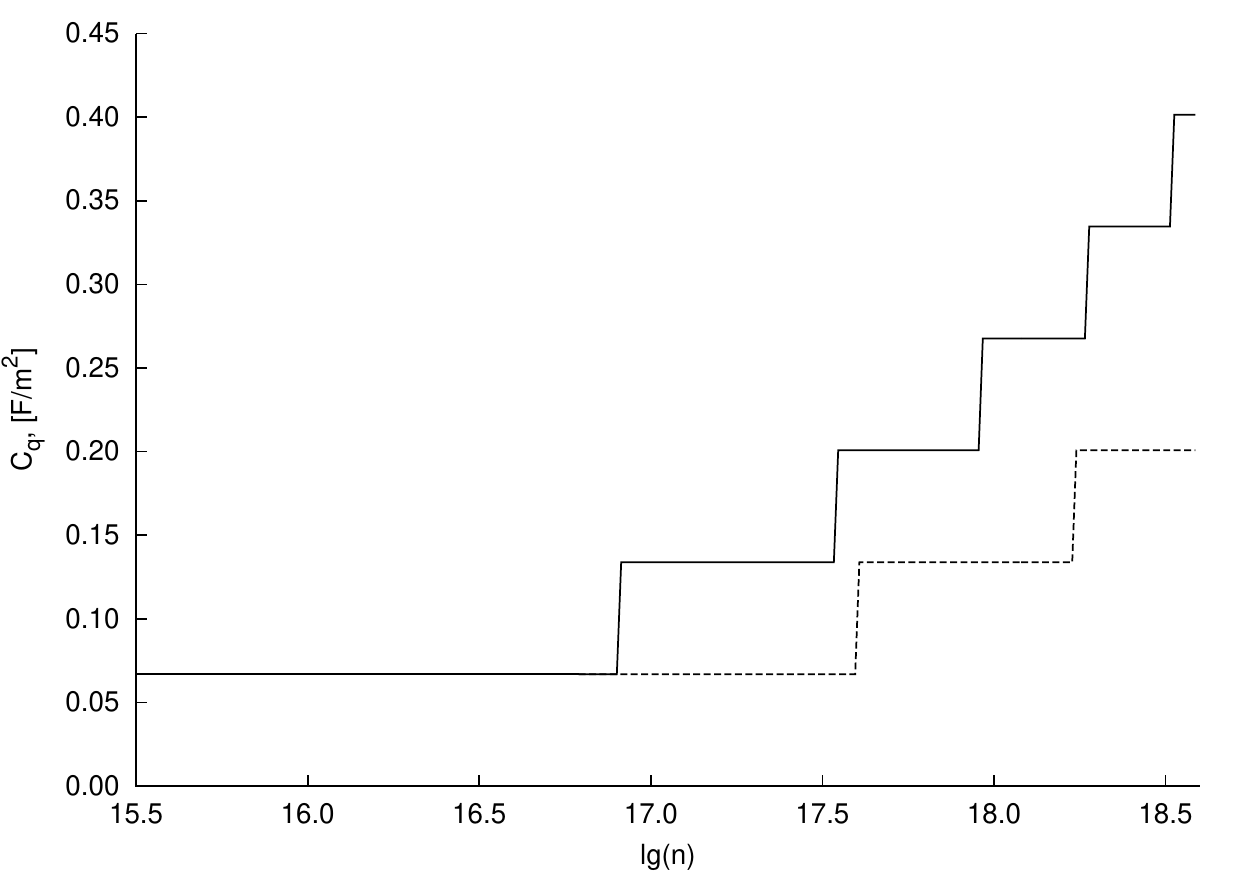}}
	\caption{\small{Dependence of a specific quantum capacity $C_q$ on $\lg(n)$ (n is the 2D concentration of electrons) for a 3D system with a potential well (in one direction) of a width $a$ with finite walls. Solid line is for $a=5$ nm, dashed line is for $a=2$ nm. An effective mass of a particle in both cases is $m^*=0.1m_0$, where $m_0$ is a ``bare'' mass of an electron.}}
	\label{fig:U1U2U3}
\end{figure}

 \newpage
 \section{Quantum capacitance for a double-well system}
 Now let's consider a symmetric rectangular double-well system of a finite depth. To find the values of eigenenergies of this system we use the obtained in \cite{Arx1:2020, Arx2:2020, Arx3:2020, Arx8:2020} expressions which allow determining energies of bound states. 
 If we put $a\rightarrow 0$ in appropriate formulas from \cite{Arx3:2020}, which means ``collapse'' of both wells then $z_b\tg (\gamma_bb)=z_a$. This corresponds to the expression for a determination of eigenenergies of a finite potential well of width $2b$. 
 
 After determination of energies of bound states of a double-well system and using both (\ref{n_mu}) and (\ref{Cq_mu}) we can find the dependence of a quantum capacitance $C_q$ on a concentration of electrons.
 The results of numerical calculations for this system are represented on Figures 4, 5.
 \begin{figure}[h!]
 	\centerline{
 		\includegraphics[clip,scale=0.9]{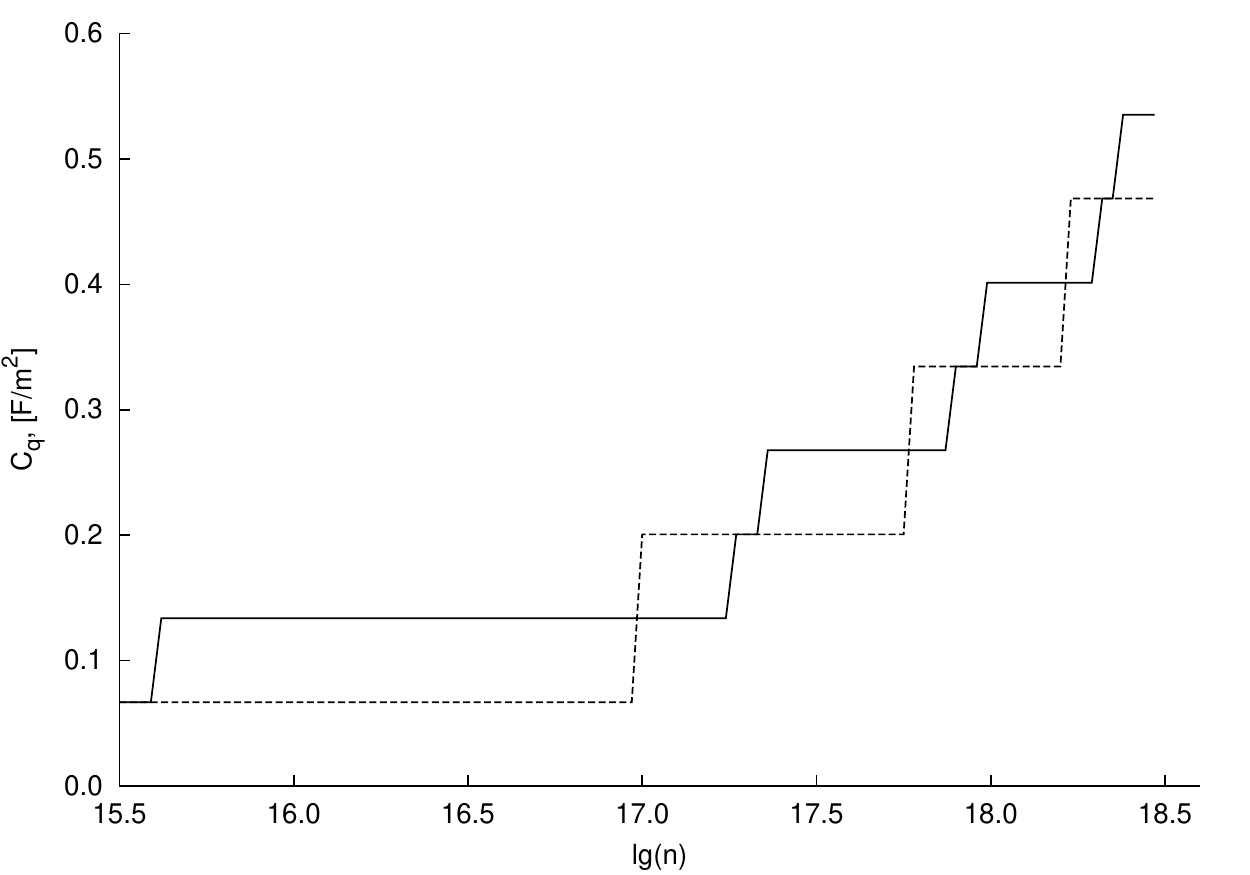}}
 	\vspace{-0.3cm}
 	\centerline{
 		\includegraphics[clip,scale=0.9]{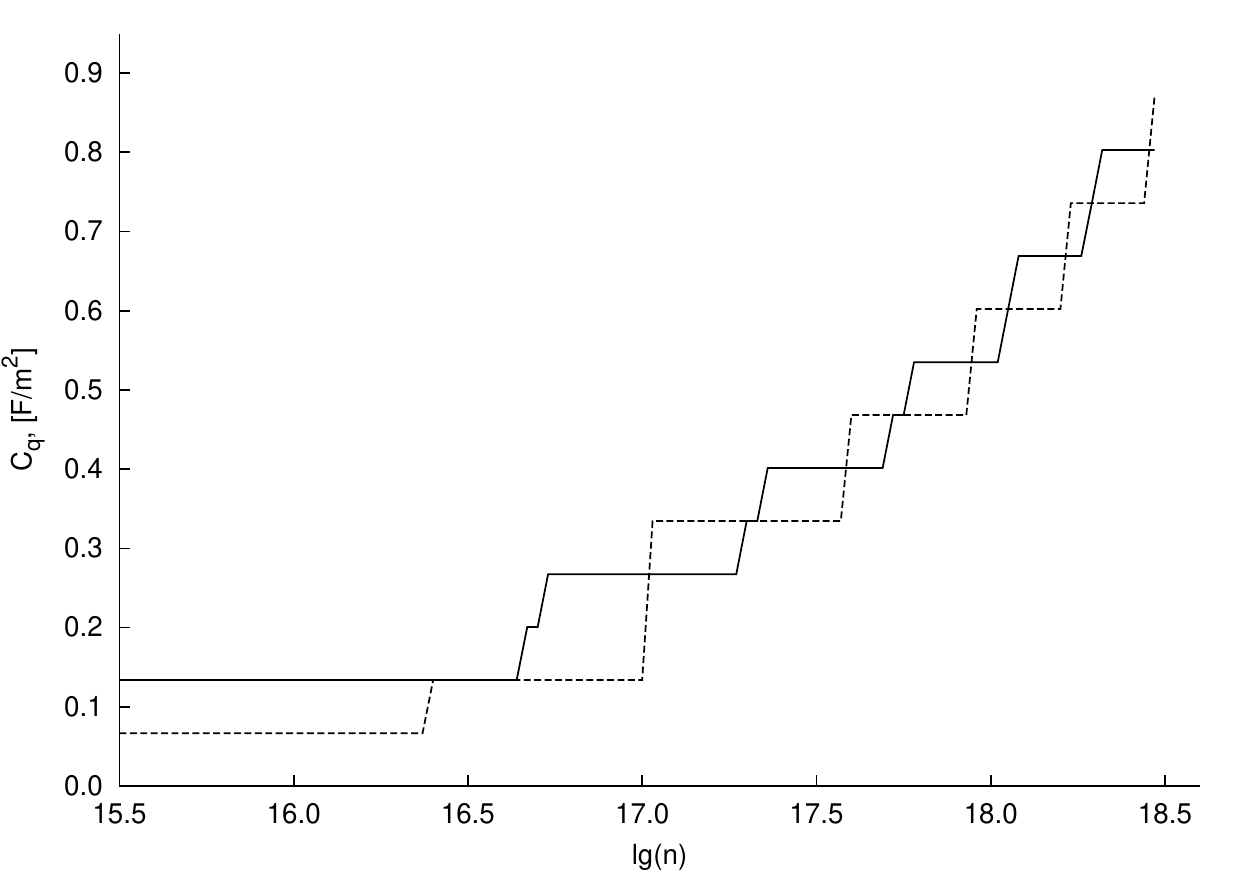}}
 	\vspace{-0.5cm}
 	\caption{\small{Dependence of a specific quantum capacity $C_q$ on $lg(n)$ (where n is a 2D concentration of electrons) for a 3D system with a double rectangular well (in one direction). The depth of each well is $10$ eV. The width of each well is $5$ nm (the upper Figure) an $10$ nm (the bottom Figure). The distance between wells is $2$ nm (solid line) and $10$ nm (dashed line).}}
 \end{figure}
 \begin{figure}[h!]
 	\centerline{
 		\includegraphics[clip,scale=0.9]{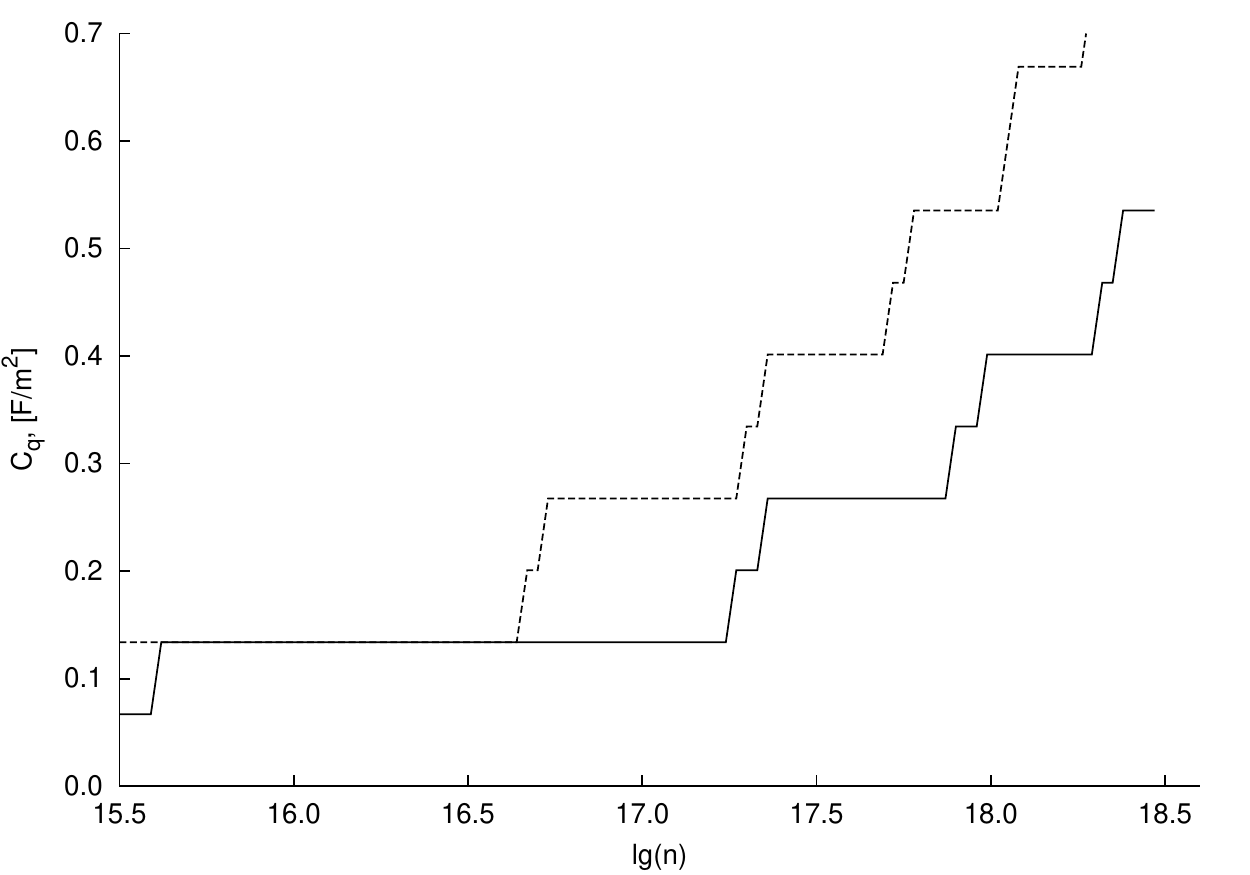}}
 	\vspace{-0.4cm}
 	\centerline{
 		\includegraphics[clip,scale=0.9]{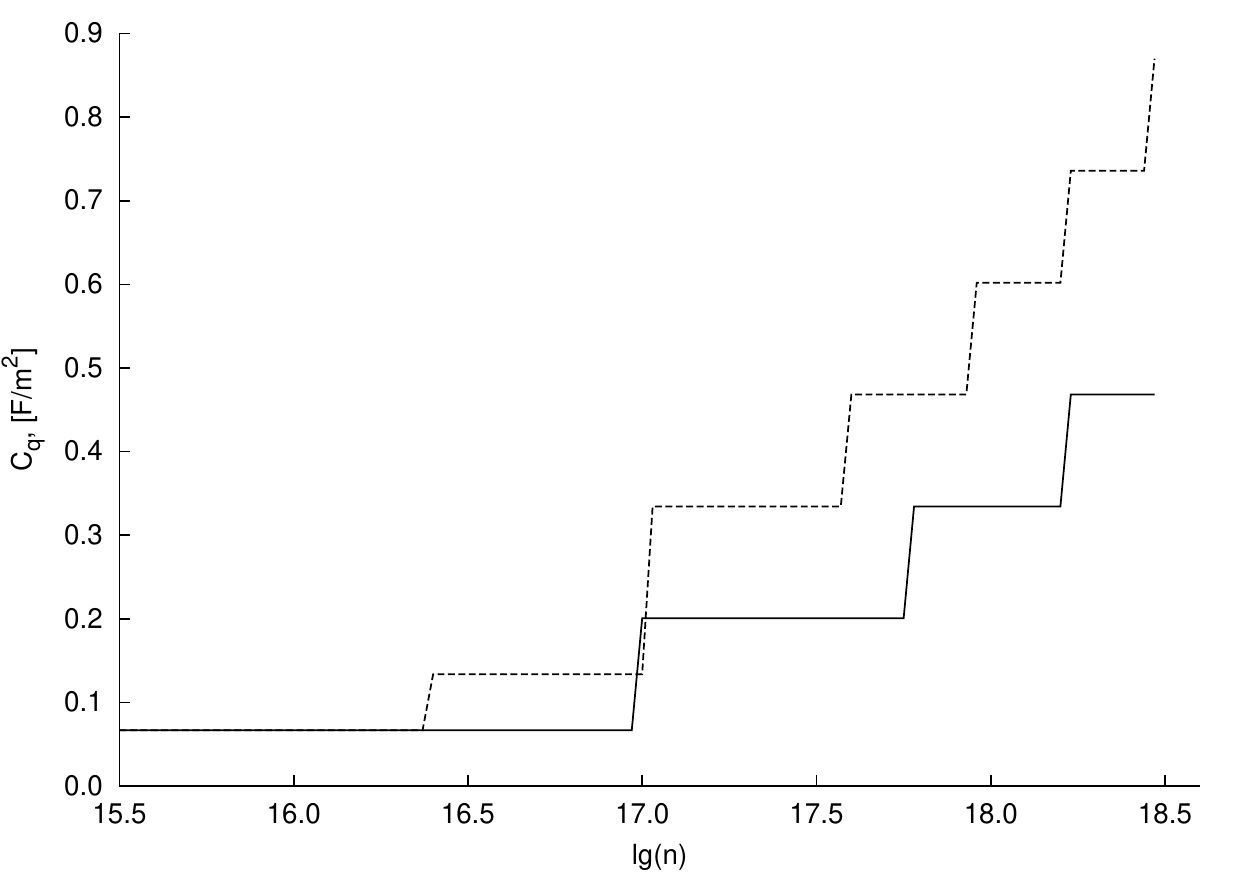}}
 	\vspace{-0.4cm}
 	\caption{\small{Dependence of a specific quantum capacity $C_q$ on $lg(n)$ (where n is a 2D concentration of electrons) for a 3D system with a double rectangular well (in one direction). The distance between wells is $2$ nm (the upper Figure) and $10$ nm (the bottom Figure). The width of each well is $5$ nm (solid line) and $10$ nm (dashed line).}}
 	\label{fig:U1U2U3}
 \end{figure}
  Now we consider a system of a parabolic double-well. We are going to calculate a quantum capacity and to compare it with the result for a rectangular double-well. 
 The potential energy in this case has the following form
 \begin{eqnarray}
 U(x)&=&ax_0^2(\theta(-x-2x_0)+\theta(x-2x_0))+a(x+x_0)^2(\theta(x+2x_0)-\theta(x))+\nonumber\\
 &+&a(x-x_0)^2(\theta(x)-\theta(x-2x_0)).
 \end{eqnarray}
 Two linearly independent solutions of a Shr\"{o}dinger equation in the first region are
 \begin{eqnarray}
 \!\!\!\!\!\!\!\!\!\psi_1(x)\!\!\!&=&\!\!\!F_{\alpha,1/2}\left(\gamma (x-x_0)^2\right)\exp\left[-\frac{1}{2}\gamma x(x-2x_0)\right],\nonumber\\
 \!\!\!\!\!\!\!\!\!\phi_1(x)\!\!\!&=&\!\!\!F_{\alpha+1/2,3/2}\left(\gamma (x\!-\!x_0)^2\right)\!(x-x_0)\exp\left[\!-\frac{1}{2}\gamma x(x\!-\!2x_0)\!\right]\!,
 \end{eqnarray} 
 where
 $\alpha=-\frac{1}{4}\left(\frac{\gamma E}{a}-1\right)$, $\gamma=\frac{\sqrt{2am}}{\hbar}$, $F_{\alpha, 1/2}(x)$ is a hypergeometric function \cite{Olver_etall:2010}. Notice that these functions in a case of $x_0=0$ coincide with appropriate Whittaker functions which we considered in \cite{Arx7:2020}.
 
 The first derivatives of $\psi_1(x)$ and $\phi_1(x)$ functions are as follows
 \begin{eqnarray}
 \psi_1'(x)&=&\left(4aF_{\alpha+1,3/2}\left(\gamma (x-x_0)^2\right)-F_{\alpha,1/2}\left(\gamma (x-x_0)^2\right)\right)\gamma(x-x_0)\exp\left[-\frac{1}{2}\gamma x(x-2x_0)\right],
 \end{eqnarray}
 \begin{eqnarray}
 \phi_1'(x)&=&\left\{\left(\frac{2}{3}(2a+1)F_{\alpha+3/2,5/2}\left(\gamma (x-x_0)^2\right)-\right.\right.\left.F_{\alpha,1/2}\left(\gamma (x-x_0)^2\right)\frac{}{}\right)-
 \gamma(x-x_0)+\nonumber\\
 &+&\left.F_{\alpha,1/2}\left(\gamma (x-x_0)^2\right)\frac{}{}\right\}\exp\left[-\frac{1}{2}\gamma x(x-2x_0)\right].
 \end{eqnarray}
 In the second region we have functions $\psi_2(x)$ and $\phi_2(x)$ which we get from $\psi_1(x)$ and $\phi_1(x)$ by changing $x_0\rightarrow -x_0$. 
  Then we calculate function $F(x_0,x_1,x_2)=\tilde{f}_1(x_0,x_1)\tilde{f}_2(x_1,x_2)$ \cite{Arx7:2020}, where $x_0=-x_0$, $x_1=0$, $x_2=x_0$ and 
 \begin{eqnarray}
 f_1(x_0,x_1)=\psi_1(x_1)\phi_1(x_0)-\psi_1(x_0)\phi_1(x_1),\nonumber\\
 f_2(x_1,x_2)=\psi_2(x_2)\phi_2(x_1)-\psi_2(x_1)\phi_2(x_2).
 \end{eqnarray}
 The relation between functions $\tilde{f}_j(x_{j-1},x_j)$ and ${f}_j(x_{j-1},x_j)$ is described in \cite{Arx7:2020}. 
 Having all these we finally determine energies of bound states and calculate a quantum capacity for this system. The numerical results are presented on the Figure 6.
 \begin{figure}[h!]
 	\centerline{
 		\includegraphics[clip,scale=0.9]{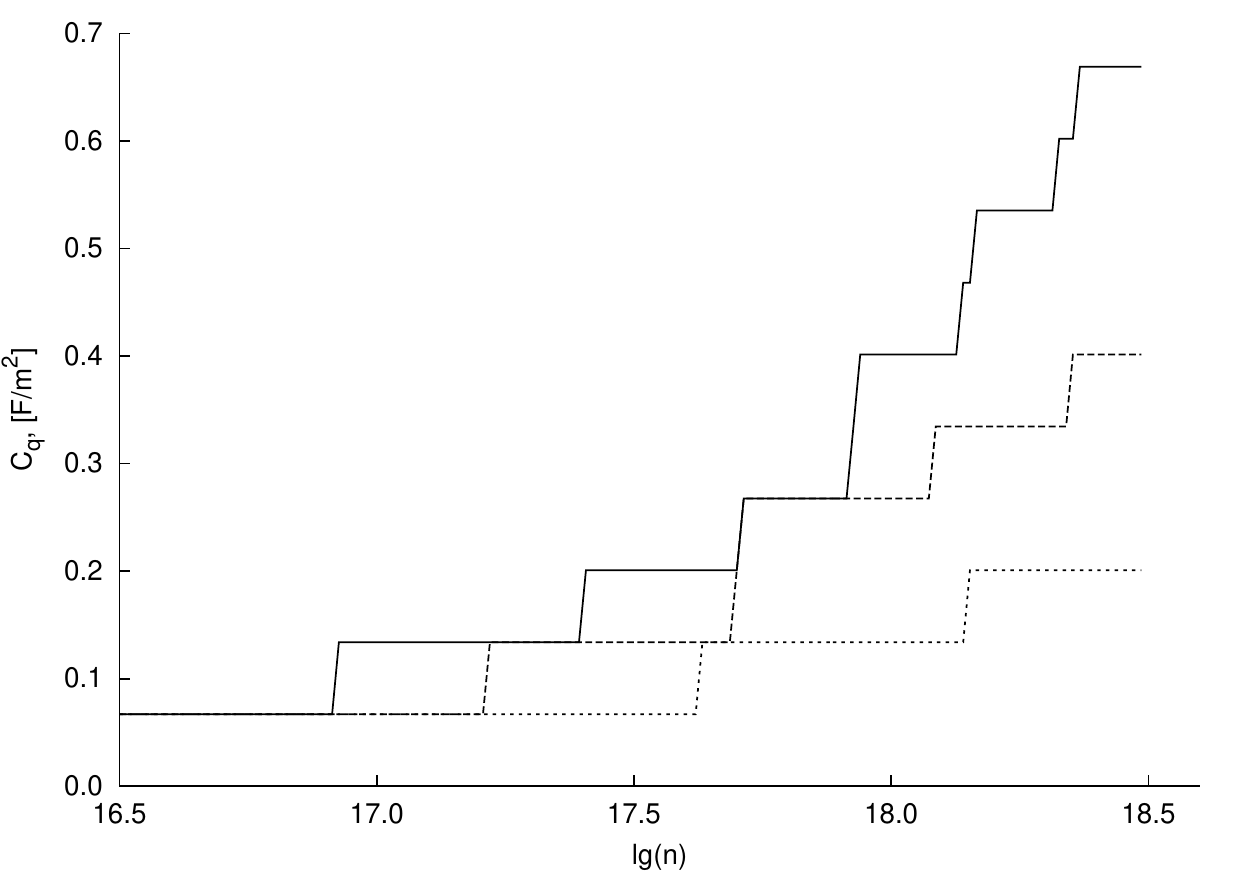}}
 	\caption{\small{Dependence of a specific quantum capacity $C_q$ on $lg(n)$ (where n is a 2D concentration of electrons) for a 3D system with a parabolic double-well (in one direction). The depth of each well is $10$ eV. Parameter $x_0=10$ nm (solid line); $x_0=5$ nm (dashed line); $x_0=2$ nm (dotted line). An effective mass of a particle is $m^*=0.1m_0$, where $m_0$ is a ``bare'' mass of an electron.}}
 	\label{fig:U1U2U3}
 \end{figure}

\section*{Conclusions}
Generally saying a quantum capacitance appears because of a low (compared to metals) density of states. This situation is a common for nanoheterostructures which are designed as a sequence of semiconductor and dielectric materials. In such systems we usually face with potential wells of different shapes which  can be approximated not   only  by rectangular wells and also by linear, parabolic, exponential and others forms for which on the base of  \cite{Arx7:2020} the solution of a Sr\"{o}dinger equation exists at least in terms of special functions. 
At the same time on the base of papers \cite{Arx4:2020, Arx5:2020, Arx6:2020, Arx9:2020} one can find the way of a quantum capacitance calculation for periodic systems and systems with zero-range singular potentials. 

Using a quantum wave impedance approach we have calculated the quantum capacitance for 2DEG in potential wells of four different forms, namely, infinite and finit rectangular well, a finit rectangular double well and a parabolic double well. We illustarted the dependence of a quantum capacitance on different parameters  of each of  these four forms of a potential. All numerical calculations were performed at temperature $T=0$. But it is no problem to provide calculations at finite temperatures. 

Having the effective method for a quantum capacitance prediction for a structures with a complicated geometry of a potential we get the key to the tailoring capacitive properties of investigated structure in aim to design nanomaterials with required characteristics. Changing different parameters of a system such we in fact modify the form of a potential well and as a result change the quantum capacitance of a system . This is important for a design of
artificial periodic structures of a nanometre size. Notice that geometric (Helmholtz) capacitance of the system is always  connected in series to the quantum capacitance.

\renewcommand\baselinestretch{1.0}\selectfont


\def\name{\vspace*{-0cm}\LARGE 
	Bibliography\thispagestyle{empty}}
\addcontentsline{toc}{chapter}{Bibliography}

{\small

	\bibliographystyle{gost780u}
	\bibliography{full.bib}
	
}

\newpage

\end{document}